\documentclass[epj]{svjour}
\pdfoutput=1
\usepackage{graphics}

\usepackage{amssymb,amsmath,rotate,color}
\begin{document}
\title{Emergence of electromotive force in precession-less rigid motion of deformed domain wall}
\author{T. Farajollahpour\inst{1,2} \and N. Darmiani\inst{1,2} \and A. Phirouznia \inst{1,2} 
}                     
\offprints{}          
\institute{Department of Physics, Azarbaijan Shahid Madani University, 53714-161, Tabriz, Iran \and Condensed Matter Computational Research Lab. Azarbaijan Shahid Madani University,
53714-161, Tabriz, Iran}
\date{Received: date / Revised version: date}
%
\abstract{
Recently it has been recognized  that the electromotive force (emf) can be induced just by the spin precession where the generation of the electromotive force has been considered as a real-space topological pumping effect. It has been shown that the amount of the electromotive force is independent of the functionality of the localized moments. It was also demonstrated that the rigid domain wall (DW) motion cannot generate electromotive force in the system. Based on real-space topological pumping approach in the current study we show that the electromotive force can be induced by rigid motion of a deformed DW. We also demonstrate that the generated electromotive force strongly depends on the DW bulging. Meanwhile results show that the DW bulging leads to generation of the electromotive force both along the axis of the DW motion and normal to the direction of motion.
} 
\maketitle
\section{Introduction}
One of the powerful techniques in condensed matter physics is 
based on the manifestation of Berry phase in electronic systems \cite{berry1984quantal,xiao2010berry,wilczek1989geometric,bohm2003geometric}. 
Berry curvature in momentum space $\Omega_{kk}$ give rise to the anomalous Hall 
effect \cite{nagaosa2010anomalous}, 
in which the Berry phase concept behave as a link between anomalous 
Hall effect and topological nature of the Hall currents. 
This effect quantitatively has been calculated by $ab initio$ methods in Refs.  \cite{turek2012ab,weischenberg2011ab,lowitzer2010coherent,wang2006ab,yao2004first}. 
As reported in Ref. \cite{neubauer2009topological} topological Hall effect is described by real space Berry curvature in 
which emergent real space Berry curvature acts like a real magnetic field. 
This effect has also been observed experimentally in several studies of this field \cite{kanazawa2011large,neubauer2009topological,muhlbauer2009skyrmion,huang2012extended}.\

In phase space, an adiabatic evolution of a wave packet generates 
various interesting Berry phase effects \cite{sundaram1999wave}. 
The evolution of the wave packet center $(r_c,k_c)$ in phase space 
is describe by a set of semi-classical equations,
\begin{align}
	\frac{\partial {\boldsymbol{{r}}_{c}}}{\partial t}=
	\frac{\partial E({\boldsymbol{{r}}_{c}},{\boldsymbol{{k}}_{c}})}
	{\hbar \partial {\boldsymbol{{k}}_{c}}}-\left( \overset{\scriptscriptstyle\leftrightarrow}
	{\boldsymbol{\Omega}}_{\boldsymbol{kr}}\cdot \frac{\partial {\boldsymbol{{r}}_{c}}}{\partial t}+\overset{\scriptscriptstyle\leftrightarrow}
	{\boldsymbol{\Omega}}_{\boldsymbol{kk}}\cdot \frac{\partial {\boldsymbol{{k}}_{c}}}
	{\partial t} \right)-{{\boldsymbol{\Omega }}_{\boldsymbol{k}t}}
\end{align}
\begin{align}
	\frac{\partial {\boldsymbol{{k}}_{c}}}{\partial t}=
	-\frac{\partial E({\boldsymbol{{r}}_{c}},{\boldsymbol{{k}}_{c}})}
	{\hbar \partial {\boldsymbol{{r}}_{c}}}+\left( \overset{\scriptscriptstyle\leftrightarrow}
	{\boldsymbol{\Omega}}_{\boldsymbol{rr}}\cdot \frac{\partial {\boldsymbol{{r}}_{c}}}{\partial t}+\overset{\scriptscriptstyle\leftrightarrow}{\boldsymbol{\Omega}}_{\boldsymbol{rk}}\cdot \frac{\partial {\boldsymbol{{k}}_{c}}}{\partial t} \right)-{{\boldsymbol{\Omega} }_{\boldsymbol{r}t}},
\end{align}
where the energy of the wave packet is $E({\boldsymbol{{r}}_{c}},\boldsymbol{k}_c)$ and
$\overset{\scriptscriptstyle\leftrightarrow}{\boldsymbol{\Omega}}$ is the 
Berry curvature tensor. The Berry curvature tensor in terms of for $(r,t)$ 
is defined by
\begin{eqnarray}
\Omega_{rt}=\frac{\partial}{\partial r}A_t-\frac{\partial}{\partial t}A_r
\end{eqnarray}
where 
\begin{eqnarray}
A_t=i\left\langle  u_k \right|\frac{\partial }{\partial t}\left| u_k \right\rangle
\end{eqnarray}
and
\begin{eqnarray}
A_r=i\left\langle  u_k \right|\frac{\partial }{\partial r}\left| u_k \right\rangle
\end{eqnarray}
known as Berry connections (gauge potentials), $\left|u_k\right\rangle$ 
is periodic part of the Bloch eigenstate of selected band. We have focused on space-time Berry curvature, $\Omega_{rt}$, 
which determines the emf of the sample. The emf in terms 
of Berry curvature is described in Refs. 
\cite{stern1992berry,barnes2007generalization,MacDonald} 
where the emf along a path $l$ can be calculated by
\begin{eqnarray}
emf={-\hbar }/{e}\int_l{dr\cdot{{\Omega }_{rt}}}.
\end{eqnarray}\

It has been shown that DWs could be driven by high electric currents. 
This effect can be explained by spin current induced spin torque which 
results in DW dynamics. In the other words DW could be driven by 
spin-transfer process originates form the spin polarized current \cite{DW_motion,DW_motion2}. 
It is interesting to know the inverse of the mentioned effect has also been predicted \cite{Berger}. 
In the inverse effect DW motion could generate an effective electromotive force. 
Generation of electromotive force (emf) by moving DW in a ferromagnetic material, 
known as a ferro-Josephson effect, which was reported in experiments performed by Yang $et al.$  \cite{yang2009universal}.
The ferro-Josephson effect has been predicted by Berger in 1986 \cite{Berger}, 
meanwhile alternative approaches reformulated this effect 
\cite{barnes2007generalization,PhysRevB.76.184434,PhysRevB.77.014409}.\

Electromotive force could be induced by Berry curvature fields in a 
time-dependent topologically nontrivial spin texture,
for example in a moving domain wall (DW). In this case it 
was shown that the space-time Berry curvature measures density 
of the emf in the real space \cite{MacDonald}. It has been 
demonstrated that the rigid motion of the DW without precession 
cannot induce electromotive force within the topological pumping approach
\cite{MacDonald}. It should be noted that the precession-less 
DW translation takes place when the DW driving magnetic field 
is less than the Walker breakdown field \cite{Walker}. Therefore 
the spin precession has a central role in generation of the 
electromotive force. This implies that even precessing of 
pinned DWs could generate electromotive force along the 
system when the magnetic field is higher than the Walker 
breakdown field \cite{MacDonald}.\

In the current work we have obtained the influence of 
the DW bulging on real space topological pumping effect 
during the rigid DW motion. Results indicate that the 
rigid motion of a deformed DW could generate an effective 
electromotive force regardless of the amount of the magnetic filed. 

If the generic Hamiltonian takes the form like 
$H={\textbf{h}}({\textbf{R}})\cdot{\widehat{\sigma}}$,
where ${{\widehat{\sigma}}}$ is the Pauli matrices, 
and ${\textbf{h}}({\textbf{R}})$ depends on parameters 
${\textbf{R}}$, then the gauge independent Berry curvature 
in parametric space ${\textbf{R}}$ is defined as \cite{nakahara2003geometry}.\

\begin{eqnarray}
\label{eqomeg}
{{\Omega }_{{{R}_{i}}{{R}_{j}}}}&=&\frac{1}{2}\frac{\partial (\beta ,\cos \gamma )}{\partial ({{R}_{i}},{{R}_{j}})}\\&=&\frac{1}{2}\det \left(
\begin{array}{cc}
\partial \beta/\partial R_1& \partial \cos \gamma/\partial R_1\\
\partial \beta/\partial R_2 & \partial \cos \gamma/\partial R_2
\end{array} \right).\nonumber
\end{eqnarray}
In which the orientation of $h(\textbf{R})$ in the parametric 
space has been a specified by the spherical angles given by 
$\cos \gamma ={{{h}_{z}}(\textbf{R})}/{| \vec{h}(\textbf{R})|},$ 
and $\beta =tan^{-1}{{{h}_{y}}(\textbf{R})}/{{{h}_{x}}(\textbf{R})}$.
In the present case in which the exchange interaction 
of the moving and localized spins is given by 
the following Hamiltonian ${{H}_{ex}}=J\hat{\sigma }.\widehat{n}(\vec{r})$ 
one can obtain $\cos \gamma=n_{z}(\theta,\phi)$ and also
$\beta =tan^{-1}\left( n_y(\theta,\phi)/ n_x(\theta,\phi)\right)$.
\\
In which the unit vector $\widehat{n}(\vec{r})=\sin \theta \cos \phi \widehat{x}+\sin \theta \sin \phi \widehat{y}+\cos \theta \widehat{z}$ describes the direction of the localized spins.

\section{Electromotive force due to the rigid domain wall motion}
The localized spin texture of the DW configuration could 
be described by the following Hamiltonian
\begin{eqnarray}
\mathcal{H}_{Loc}&=&-J_{loc}\sum_{<i,j>}\vec{n}(r_i)\cdot \vec{n}(r_j)-\sum_{i}\gamma_z \vec{n}_m\cdot\vec{n}(r_i)\nonumber\\&&-\sum_i (K_{\parallel}\vec{n}(r_i)\cdot \hat{e}-K_{\perp}\vec{n}(r_i)\cdot \hat{e}')+\mathcal{V}_{pin}
\end{eqnarray}
In which the first term describes the Heisenberg 
Hamiltonian of the first neighbor spins where $J_{loc}$ 
denotes the exchange coupling constant between the 
localized spins $e$ denotes the easy axis while $e'$ 
being the direction of the hard axis. $K_{\parallel}$, 
$K_{\perp}$ stand for anisotropic constants for easy 
and hard axis respectively.  $n(r_i)$ describes the 
direction of the localized spins as mentioned before.  
$\gamma_z$ is the Zeeman splitting, $\mathcal{V}_{pin}$ 
stands for pinning potential and $\vec{n}_m$ denotes 
the unit vector describing the direction of the external 
magnetic field. The competition between the exchange 
and anisotropic couplings determines the DW width 
which is given by $d\sim\sqrt{K_{\parallel}/J_{loc}}$.\

Meanwhile the dynamics of the local spins have been 
determined with the Landau and Lifshitz equation which 
could be expressed as
\begin{eqnarray}
\frac{d\vec{n}(r_i,t)}{dt}=\gamma \vec{n}(r_i,t)\times \vec{n}_m+\alpha \vec{n}(r_i,t)\times\frac{d\vec{n}(r_i,t)}{dt},
\end{eqnarray}
where $\alpha$ is the damping constant.\

DW motion can be considered as a successive spin 
switching of the local precessing moments as a 
result of the damping procedure. When the magnetic 
field is higher than depinning field spin falls 
along the direction of effective field during the 
damping process. This was known as spin switching process.\

DW motion itself could take place in different regimes. 
When the magnetic field is less than the Walker breakdown 
field \cite{Walker} DW moves actually in a shape preserved 
manner. In this case DW makes a rigid movement since the 
spin switching process takes place almost abruptly. 
Therefore rigid DW motion rests largely on how fast 
the spin switching takes place. On the other hand when 
the magnetic field is higher than Walker breakdown 
field, DW will precess during its translational motion. 
As shown by Yang and et al for a moving DW without 
precession i.e for rigid DW motion there is no 
electromotive force in the system \cite{MacDonald}.
This could be demonstrated for both Bloch and N\'{e}el type DWs. 
In the absence of the precession, when the spin switching takes 
place abruptly, rigid DW motion could be described by the 
following profiles
\begin{eqnarray}
\phi&=&const \nonumber\\
\theta(x,t)&&=\pi/2+(\pi/2)\tanh((x-v_Dt)/d)
\end{eqnarray}
for Bloch type DWs and
\begin{eqnarray}
\phi(x,t)&=&(\pi/2)\tanh((x-v_Dt)/d)\nonumber\\
 \theta&&=const
\end{eqnarray}
for N\'{e}el type DWs.\

The space and time dependence of $\beta$ and $\cos\gamma$ 
are just given by the spherical angles ($\theta$ and $\phi$).
Therefore it can be easily shown that
\begin{eqnarray}
{{\Omega }_{{x}{t}}}&=&\frac{1}{2}\det \left(
\begin{array}{cc}
\partial \beta/\partial x& \partial \cos \gamma/\partial x\\
\partial \beta/\partial t & \partial \cos \gamma/\partial t
\end{array} \right)\nonumber\\
&=&\frac{1}{2}\det \left(
\begin{array}{cc}
\partial_x \theta& \partial_x \phi\\
\partial_t \theta & \partial_t \phi
\end{array} \right)\det \left(
\begin{array}{cc}
{\partial_\theta \beta}& {\partial_\theta \cos \gamma}\\
{\partial_\phi \beta} & {\partial_\phi \cos \gamma}
\end{array} \right)\nonumber\\
&=&\Omega_{\theta \phi} \det \left(
\begin{array}{cc}
\partial \theta/\partial x& \partial \phi/\partial x\\
\partial \theta/\partial t & \partial \phi/\partial t
\end{array} \right). \label{iden}
\end{eqnarray}\

Using the Eq.~\ref{iden} it can be inferred that 
electromotive force generated by DW motion identically 
vanishes for rigid translation of the DW in both of 
the N\'{e}el and Bloch-type configurations. This 
can be easily obtained if we consider that for 
either of the profiles given for the rigid motion of the DW  we have
\begin{eqnarray}\det \left(
\begin{array}{cc}
\partial \theta/\partial x& \partial \phi/\partial x\\
\partial \theta/\partial t & \partial \phi/\partial t
\end{array} \right)&=&\frac{\partial (\theta ,\phi )}{\partial ({{x ,\ t}})}\\&=&0\nonumber
\end{eqnarray}
and therefore ${{\Omega }_{{x}{t}}}=0$.
\section{Electromotive force generated by the rigid motion of a deformed domain wall}
If we consider the bulging effect in the DW it could 
be demonstrated that the rigid motion of the DW 
generates an effective electromotive force inside the system.
DW bulging could be characterizes by a position 
dependent rotation axis $\hat{n}_D(\vec{r})$ 
and the boundary curve of the DW, where we have 
assumed that the DW rotation axis should be 
directed vertically to the DW boundary surface. 
This means that when the DW boundary has been specified 
by a surface described by $F(r_0)=c$ then DW 
rotation axis is given by $\hat{n}_D(\vec{r})=\nabla F/|\nabla F|$ 
at each point of the boundary curve.\

In the current case we have assumed a deformed 
Bloch type DW. If the direction of the local 
magnetization is upward at the left boundary 
of the DW then local direction of the magnetization 
could be obtained by the spin rotation operator as follows
\begin{eqnarray}
\sigma\cdot \vec{n}&=&R^{\dag}_{(\theta)}\sigma_z R_{(\theta)}\nonumber\\
&=&\cos^2(\theta(X_D)/2)\sigma_z\nonumber\\&&+\sin^2(\theta(X_D)/2)\cos^2(\theta(X_D)/2)[\sigma\cdot \vec{n}_D,\sigma_z]\nonumber\\&&+\sin^2(\theta(X_D)/2)(\sigma\cdot \vec{n}_D)\sigma_z(\sigma\cdot \vec{n}),
\end{eqnarray}
in which
\begin{eqnarray}
R_{\theta(X_D)}&=&\exp(-i \sigma\cdot \vec{n}_D(\vec{r})\theta(X_D))\\
&=&1\cos(\theta(X_D)/2)-i \sigma\cdot \vec{n}_D(\vec{r})\sin(\theta(X_D)/2),\nonumber
\end{eqnarray}
$R_{\theta}$ is the spin rotation operator and 
$X_D(r,t)=\hat{n}_D(r)\cdot \vec{r}(t)$ is the 
projection of position vector along the direction 
of the DW rotation axis. Accordingly we can obtain
\begin{eqnarray}
\sigma\cdot \vec{n}&=&(\cos(\theta(X_D))+2n^2_{Dz}\sin^2(\theta(X_D)/2))\sigma_z\\&&+(2n_{Dx}n_{Dz}\sin^2(\theta(X_D)/2)-n_{Dy}\sin(\theta(X_D)))\sigma_x\nonumber\\
&&+(2n_{Dy}n_{Dz}\sin^2(\theta(X_D)/2)+n_{Dx}\sin(\theta(X_D)))\sigma_y.\nonumber
\end{eqnarray}
Therefore the unit vector of the local magnetization could be extracted as follows
\begin{eqnarray}
\vec{n}(\vec{r})&&=\\
&&\left(
\begin{array}{c}
\cos(\theta(X_D)+2n^2_{Dz}\sin^2(\theta(X_D)/2))\\
2n_{Dx}n_{Dz}\sin^2(\theta(X_D)/2)-n_{Dy}\sin(\theta(X_D))\\
2n_{Dy}n_{Dz}\sin^2(\theta(X_D)/2)+n_{Dx}\sin(\theta(X_D))
\end{array} \right).\nonumber
\end{eqnarray}
\begin{figure}
\resizebox{0.5\textwidth}{!}{%
  \includegraphics{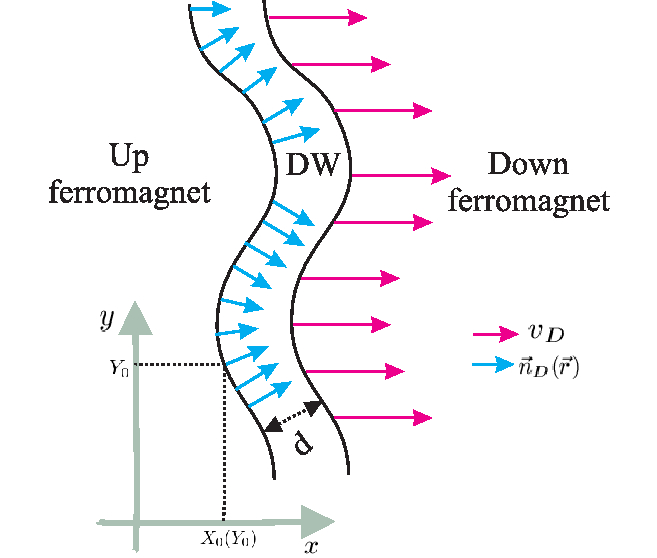}
}
\caption{A typical form of the deformed DW in the real space. 
	DW has been placed between two oppositely directed magnetic 
	domains. A Bloch-type DW has been considered in which the 
	DW rotation axis located in the plane of the system. The 
	bulging of the DW can be formulated if we consider that 
	the DW rotation axis is always vertically oriented to 
	the DW boundaries. $n_D$ denotes the DW rotation axis 
	while $v_D$ indicates the speed of the DW and $d$ is 
	the DW width. }
\label{bulging}
\end{figure}
\\
If we restrict ourself to the case of the two-dimensional 
DWs i.e. when ($n_{Dz}=0$) we obtain the simplest case of 
two dimensional bulging as follows
\begin{eqnarray}
\vec{n}(\vec{r})&=&-n_{Dy}\sin(\theta(X_D))\hat{x}\nonumber\\&&+n_{Dx}\sin(\theta(X_D))\hat{y}+\cos(\theta(X_D))\hat{z}.
\end{eqnarray}
DW width has been assumed to be locally constant during 
the DW rigid motion. The DW motion has been considered 
as a successive switching of the local spins that as 
mentioned before. In the case of the rigid DW motion 
it was assumed that the spin switching time is very 
short and a localized spin switches abruptly without 
precession. Meanwhile it was assumed that the DW 
rigidly moves along the $x$ axis. 
Since in this case we can write
\begin{eqnarray}
\phi(y)=tan^{-1}(-n_{Dy}(y)/n_{Dx}(y))=-\phi_D
\end{eqnarray}
and
\begin{eqnarray}
X_D(r,t)&=&\vec{n}_{D}\cdot\vec{r}
\\&=&n_{Dx}(x-X_0(Y_0)-v_Dt)+n_{Dy}(y-Y_0)\nonumber
\end{eqnarray}
where $X_0(Y_0)$ is the left boundary of the DW 
at the initial time. This curved boundary preserves 
its shape during the DW motion which sweeps and 
removes the downward ferromagnetic region (right 
region in Fig.~\ref{bulging}) during the motion of the DW.\

The space time Berry curvature is given by
\begin{eqnarray}
\Omega_{yt}&=&\Omega_{\theta\phi}(\partial_y \phi)(\partial_t \theta(X_D))\\&=&\Omega_{\theta\phi}n_{Dx}v_D\partial_{X_D}\theta(X_D))(n_{Dx}\partial_yn_{Dy}-n_{Dy}\partial_yn_{Dx}).\nonumber
\end{eqnarray}
Since we can write $n^2_{Dx}+n^2_{Dy}=1$, 
therefore 
\begin{eqnarray}
n_{Dx}(n_{Dx}\partial_yn_{Dy}-n_{Dy}\partial_yn_{Dx})=\partial_yn_{Dy}
\end{eqnarray} 
Meanwhile in the absence of the Rashba interaction we have $\beta=\phi$ and $\cos \gamma= \cos \theta$ and therefore 
$\Omega_{\theta\phi}=1/2\sin(\theta)$. Accordingly
\begin{eqnarray}
\Omega_{yt}=\frac{\pi}{4d}\sin(\theta(X_D))v_D\cosh^{-2}(X_D/d)\partial_y n_{Dy}
\end{eqnarray}
and similarly we can obtain
\begin{eqnarray}
\Omega_{xt}=\frac{\pi}{4d}\sin(\theta(X_D))v_D\cosh^{-2}(X_D/d)\partial_x n_{Dy}.
\end{eqnarray}
Unlike the straight DWs in which the electromotive force 
could only generated as a result of the precession, 
regardless of the DW motion, in the current case for 
deformed DWs we have shown that the electromotive force 
could be generated along the direction of motion and even 
normal to this direction.\

It should be considered that the averaged electromotive 
force along the direction of the DW motion could vanish 
identically when the DW curve has mirror symmetry and 
the direction of the DW speed is parallel to the plane 
of mirror symmetry. Meanwhile, even in this case the 
normal electromotive force still survives and the rigid 
DW motion could generate an electromotive force normal 
to the direction of motion i.e. perpendicular to the 
mirror symmetry plane. Therefore the electromotive 
force induced by the rigid motion of the domain wall 
can identically vanish when the domain wall has two 
vertical mirror symmetry planes. However in this case 
the domain wall should be formed as a closed symmetric 
curve i.e. one of the magnetic domains should be 
surrounded by its opposite counterpart.\

Without loss of generality we can assume that the 
domain wall motion is along the x direction.
If we assume that the $xz$ plane is the mirror 
plane of the DW therefore $X_0(-Y_0)=X_0(Y_0)$, 
$n_{Dx}(x,-y)=n_{Dx}(x,y)$ and $n_{Dy}(x,-y)=-n_{Dy}(x,y)$. 
Therefore $X_D$ remains unchanged under the 
coordinate change: $y\rightarrow-y$ and $Y_0\rightarrow-Y_0$. 
In this case since $\Omega_{xt}$ is an odd function of $y$ 
the spatial average of the electromotive force along the 
direction of the DW motion, $(-\hbar/eL)\int_S dx dy \Omega_{xt}$, 
vanishes identically. Where the integration has been performed 
over the DW area which has been specified by $S$ and $L$ is 
the length of the DW. Meanwhile in the normal direction 
$\Omega_{yt}$ is an even function and the averaged 
electromotive force, $(-\hbar/eL)\int_S dx dy \Omega_{yt}$, 
cannot vanish simultaneously. Finally we can conclude that 
electromotive force induced by the rigid DW motion cannot 
go all the way to zero for any arbitrary open curved DWs.\

It should be noted that in the previous studies in this filed  \cite{Volovik,PhysRevLett.102.086601,PhysRevLett.107.236602} spin-motive force was shown to be proportional to $|\partial \textbf{m}/\partial t \times \partial \textbf{m}/\partial x|$. Since in the precession-less limit $\partial\textbf{m}/\partial t$ identically vanishes, this means that the zeroth order spin-motive force is zero for rigid motion of the DW. Meanwhile if we consider the perturbative induced effects in the effective field of the system, perturbative terms of $\partial\textbf{m}/\partial t$ could be non-zero and accountable. Spatially non-homogeneous magnetization results in effective magnetic field on conduction electrons given by $\textbf{B}_i\sim\epsilon^{ijk}(\partial_j \textbf{m}\times \partial_k \textbf{m})\cdot\textbf{m}$.
Since the spin torque on the local magnetization is the opposite of the torque exerted on the conduction electrons. Therefore local magnetization even in the precession-less limit has time evolution
\cite{PhysRevLett.102.086601}.

\section{Conclusion}
In the current study we have developed the Berry 
curvature based theory of the DW electromotive 
force for the deformed DWs. In the previous 
studies it has been demonstrated that for 
straight DWs rigid motion of the DW could 
not contribute in the ferro-Josephson effect 
\cite{MacDonald}. In the other words this type 
of domain wall dynamics which takes place below 
the Walker field could not generate electromotive 
force and the spin precession plays the central 
role in the electromotive force generation \cite{MacDonald}. 
However in the current study we have demonstrated 
that this is not the case for the deformed DWs. 
It has been shown that the DW bulging results 
in effective electromotive force in the case 
of rigid the motion of the domain wall. In 
addition we have shown that the generated 
electromotive force could be either along 
the DW motion axis or in the transverse direction.

\section*{Acknowledgment}
This research has been supported by Azarbaijan Shahid Madani university.
\bibliographystyle{spphys}
\bibliography{refnew}
%

\end{document}